\documentclass[showpacs]{revtex4}
\input epsf
\def\be{\begin{equation}}
\def\ee{\end{equation}}
\def\bea{\begin{eqnarray}}
\def\eea{\end{eqnarray}}
\def\lsim{\mathrel{\rlap{\lower4pt\hbox{\hskip1pt$\sim$}}
    \raise1pt\hbox{$<$}}}
\def\gsim{\mathrel{\rlap{\lower4pt\hbox{\hskip1pt$\sim$}}
    \raise1pt\hbox{$>$}}}
\def\sa{\sigma_8}
\def\omd{\Omega^d}
\def\wda{\omega^d}
\newcommand{\omdn}{\Omega ^{d}_0}
\newcommand{\omdsf}{\bar{\Omega} _{\rm d} ^{\rm sf}}
\newcommand{\aeq}{a_{\rm eq}}

\begin{document}

\title{WMAP constraints on Cardassian model}

\author{A.A.Sen$^1${\footnote{anjan@x9.ist.utl.pt}
and S.Sen$^2${\footnote{somasri@cosmo.fis.fc.ul.pt}}}}

\affiliation{$^1$Centro Multidisciplinar de Astrof\'{\i}sica,
 Departamento de F\'\i sica, Instituto Superior T\'ecnico \\
 Av. Rovisco Pais 1, 1049-001 Lisboa, Portugal}

\vskip 0.2cm

\affiliation{$^2$CAAUL,
 Departamento de F\'\i sica da FCUL, Campo Grande, 1749-016 Lisboa, Portugal.}

\date{\today}

\begin{abstract}
We have investigated the constraints on the Cardassian model using the recent results from the Wilkinson Microwave Anisotropy Probe (WMAP) for the locations of the peaks of the Cosmic Microwave Background (CMB) anistropy spectrum. We find that the model is consistent with the recent observational data for certain  range of the model parameter $n$ and the cosmological parameters. We find that the Cardassian model is favoured compared to $\Lambda$CDM model, for a higher spectral index ($n_s \approx 1$) together with lower value of Hubble parameter $h$ ($h\leq 0.71$). But for smaller values of $n_s$, both $\Lambda$CDM and Cardassian are equally favoured. Also, irrespective of Supernova constraint, CMB data alone predicts current acceleration of the universe in this model. We have also studied the constraint on the $\sa$, the rms density fluctuations at $8h^{-1}$ Mpc scale.  
\end{abstract}
\maketitle
\vskip 5mm

\section{Introduction}
The recent era has been dedicated to precision cosmology. Different 
remarkable observations with great precision has set new direction 
in realising the details of the universal evolution. Over the last few 
years we have developed a firm broad outline about our universe to be 
spatially flat, homogeneous, isotropic on large scales and composing of 
radiation, ordinary matter including baryon and neutrinos, cold dark 
matter and a huge share of dark energy. There have been a host of 
phenomenological models in support of this picture, which sets strong 
limits on the cosmological parameters. But the latest 
Wilkinson Microwave Anisotropy Probe (WMAP)\cite{wmap} has set a new challenge
 for these models. WMAP has shown incredible precision in predicting different 
cosmological parameters including the positions of the peaks in CMB 
anisotropic spectrum. WAMP's accurate determination of angular power spectrum  
along with other astronomical data sets like SNIa observations\cite{sn1,sn2} 
and 2dF galaxy survey\cite{2df}, put significant limit on the cosmological 
parameters, which should be defended by the various phenomenological models.

Since the significant observations regarding the SNIa \cite{sn1,sn2},  
several models both simple and complicated ones,  have been used so far to justify 
the late time accelerating universe and at the same time proposing different ranges for the cosmological 
parameters. With $\Lambda$CDM among the simple and Quintessence\cite{quint}, 
K-essence\cite{ksen} 
and several others\cite{others} as more complicated ones, each one aims 
in describing the universe in accordance to more precised observational 
agreement.     
One more thing common here is that these models accepts the idea of dark 
energy in any form of field or fluid. While there have been many other 
interesting ideas like the generalised Chaplygin gas\cite{chap} or 
rolling tachyons\cite{paddy}, a 
different proposal has been put by Freese and Lewis\cite{freese}. 
In this new approach 
the new cosmology is obtained by modifying the dynamics of the universe, 
namely, the Friedman equation, without seeking to another unknown component 
like dark energy.  

They have proposed an universe composed of only 
radiation and matter(including baryon and cold dark matter) which expands 
in the speed up fashion if an empirical term, called Cardassian term, 
is added to the Friedman equation.
\be
H^2=A~\rho+B~\rho^n,
\ee
where $A={8\pi G\over{3}}$ and $B$ and $n$ are constants and are the 
parameters of the model. Here the energy density $(\rho)$ contains only matter 
$(\rho_m)$ and 
radiation $(\rho_r)$, i.e, $\rho=\rho_m+\rho_r$. 
Since at present 
$\rho_m>>\rho_r$, $\rho$ can be considered consisting of $\rho_m$ only at present. 
The new term, dominates only recently at redshift $\sim 1$. 
To provide the requisite acceleration of the universe 
as the outcome of the dominance of this term, $n$ should be $<2/3$.
The model has two main parameters $n$ and $B$.

There are several interpretations for the origin of this new ``Cardassian 
term'' appearing in the Einstein's equation (1)\cite{motiv}.
To accommodate flatness with only matter and radiation in this new picture, 
the critical energy density has been modified and expressed as a function of the original one. Here the total density parameter at present $\Omega_0$ is 
defined as $\frac{\rho_{m0}+\rho_{r0}}{\rho_c}$ where the redefined critical energy density $\rho_c$  is related to the actual one $\rho_{ca}(=3H_0^2/8\pi G)$ as,
\be
\rho_c=\rho_{ca}\times F(n,B)
\ee
where
\bea
F(n,B)&=&[1+\frac{B}{A}\rho_{m0}^{n-1}(1+\frac{\Omega_{r0}}{\Omega_{m0}})^{n-1} ]^{-1}
\eea
$\Omega_{m0}$ and $\Omega_{r0}$ are two parameters  defined as 
$\Omega_{m0}=\frac{\rho_{mo}}{\rho_{ca}}$ and $\Omega_{r0}=\frac{\rho_{ro}}{\rho_{ca}}$ 
respectively.

Now, as matter and radiation evolves, equation (1) can be expressed as 
the following,
\be
H^2=A~[\rho_{m0}a^{-3}(1+\frac{\Omega_{r0}}{\Omega_{m0}}a^{-1})+\frac{B}{A}
~\rho_{m0}^n~a^{-3n}(1+\frac{\Omega_{r0}}{\Omega_{m0}}a^{-1})^n]
\ee
From equation (3) it is very straight forward to express B in terms of 
$\Omega_{r0}$ and $\Omega_{m0}$ to be
\be
\frac{B}{A}\rho_{m0}^{n-1}=(\frac{1-\Omega_{r0}-\Omega_{m0}}{\Omega_{m0}})(1+\frac{\Omega_{r0}}{\Omega_{m0}})^{-n}.
\ee
Substituting this expression in equation (4), 
equation (1) is finally recast in the following fashion 
\be
H^2=\Omega_{m0}H^2_0 a^{-4}~\left[(a+\frac{\Omega_{r0}}{\Omega_{m0}})+a^{-4n+4}
\left(\frac{1-\Omega_{r0}-\Omega_{m0}}{\Omega_{m0}}\right)
\left(\frac{a+\frac{\Omega_{r0}}{\Omega_{m0}}}{1+\frac{\Omega_{r0}}{\Omega_{m0}}}\right)^n\right]
\ee

One should note that the only model parameter appearing in the above 
expression is $n$. (For details discussion please refer \cite{our})
The other important point to notice is that the value $n=0$ corresponds 
to a $\Lambda$CDM model. This will be crucial when we shall talk later 
about the allowed region of the parameter space.

\begin{figure}[t]
\centering
\leavevmode \epsfysize=12cm \epsfbox{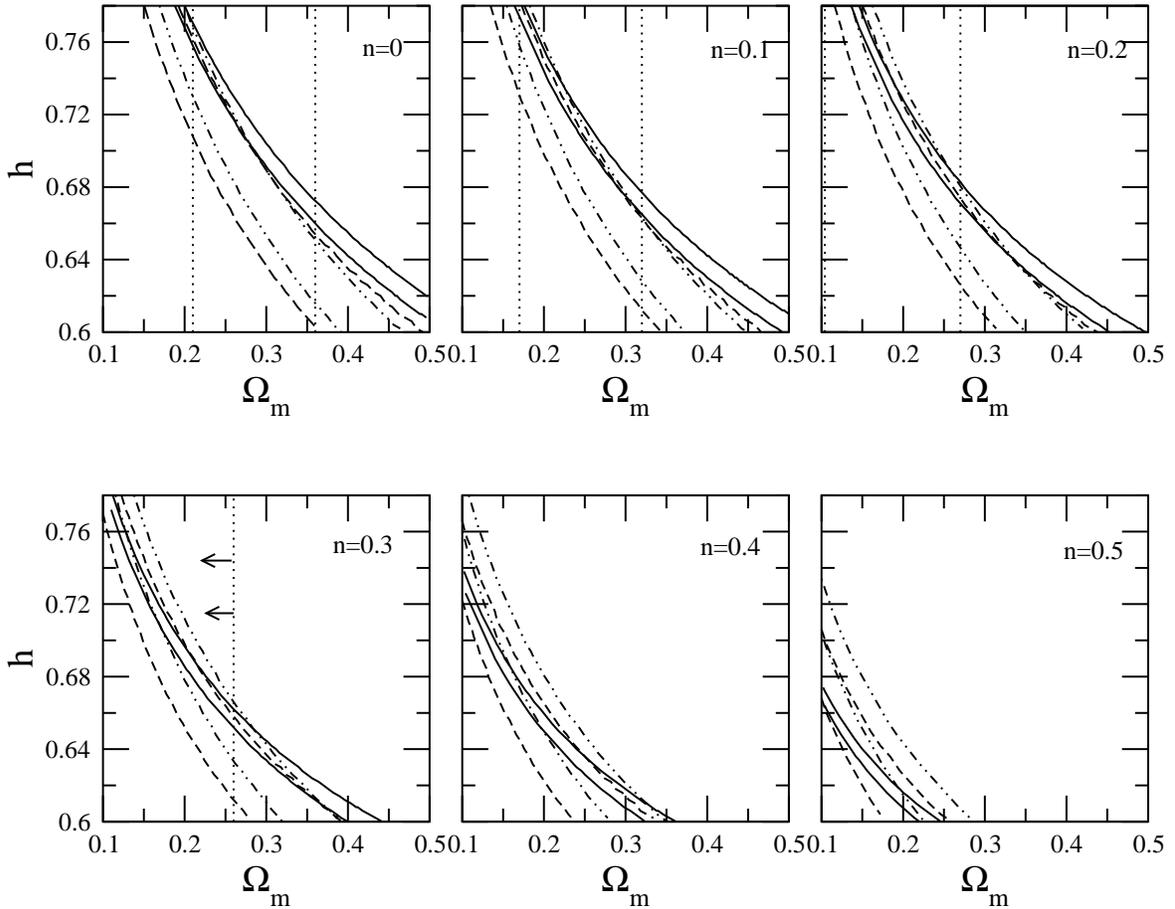}\\
\vskip 0.1cm
\caption{Contour plots of the first, second and third Doppler peak locations in the ($\Omega_{m},h$)
 plane for different values of $n$ and $n_{s}=1$. Full and dashed contours correspond to the WMAP 
bound of first and second peak locations and the dot-dashed contour corresponds to the Boomerang 
bound on third peak location. The region within the dotted line corresponds to the $1-\sigma$ 
confidence region for $\Omega_{m}$ for the Supernova data. The allowed region is the intersection of
 all the contours.}
\label{fig:figure1}
\end{figure}

In one of our very recent papers\cite{our} we have considered the constraints imposed on this model's parameters from the CMB measurements like BOOMERanG and Archeop and the Supernova results. We found the model as a quite aspiring as an alternative to dark energy models with interesting limits on its parameter space (For other investigations in this regard please refer to \cite{other}). Now in the light of the very precised measurements of WMAP it is certainly very much worthy to verify the consistency of the model and found the new restricted parameter space. We further study the constraints on $\sigma_8$ in our model.

\section{The constraints from WMAP}
The CMBR peaks arise from oscillation of the primeval plasma just before 
the universe becomes translucent. The oscillation of the tightly bound 
photon-baryon fluid is a result of the balance between the gravitational 
interaction and photon pressure and this oscillations gives rise to the 
peaks and troughs in the temperature anisotropic spectrum. Locations of 
the peaks at different angular momentum $l$, resulted from this oscillation, 
 depend on the acoustic scale $l_A$ which in turn is related to the 
angular diameter distance to the last scattering $D$ and the sound horizon 
at the last scattering $s_{ls}$ as $\frac{\pi D}{s_{ls}}$\cite{hu}. 
To a good approximation this ratio for $l_A$ can also be represented by 
the simple expression\cite{doran1} 
\be  
l_A=\pi\frac{\tilde\tau_0-\tilde\tau_{ls}}{\bar{c_s}\tilde\tau_{ls}}.
\ee 
where $\tilde\tau(=\int a^{-1}dt)$ is the conformal time and the subscript $0$ 
and $ls$ represent the time at present and at the last scattering era 
respectively. $\bar{c_s}$ is the average sound speed before 
last scattering defined as
\be
\bar{c_s}\equiv \tilde\tau_{ls}^{-1}\int_0^{\tilde\tau_{ls}} c_s d\tilde\tau,
\ee
with 
\be
c_s^{-2}=3+\frac{9}{4}~\frac{\rho_b}{\rho_r}
\ee
$\frac{\rho_b}{\rho_r}$ is the ratio of baryon to photon energy density. 

\begin{figure}[t]
\centering
\leavevmode \epsfysize=12cm \epsfbox{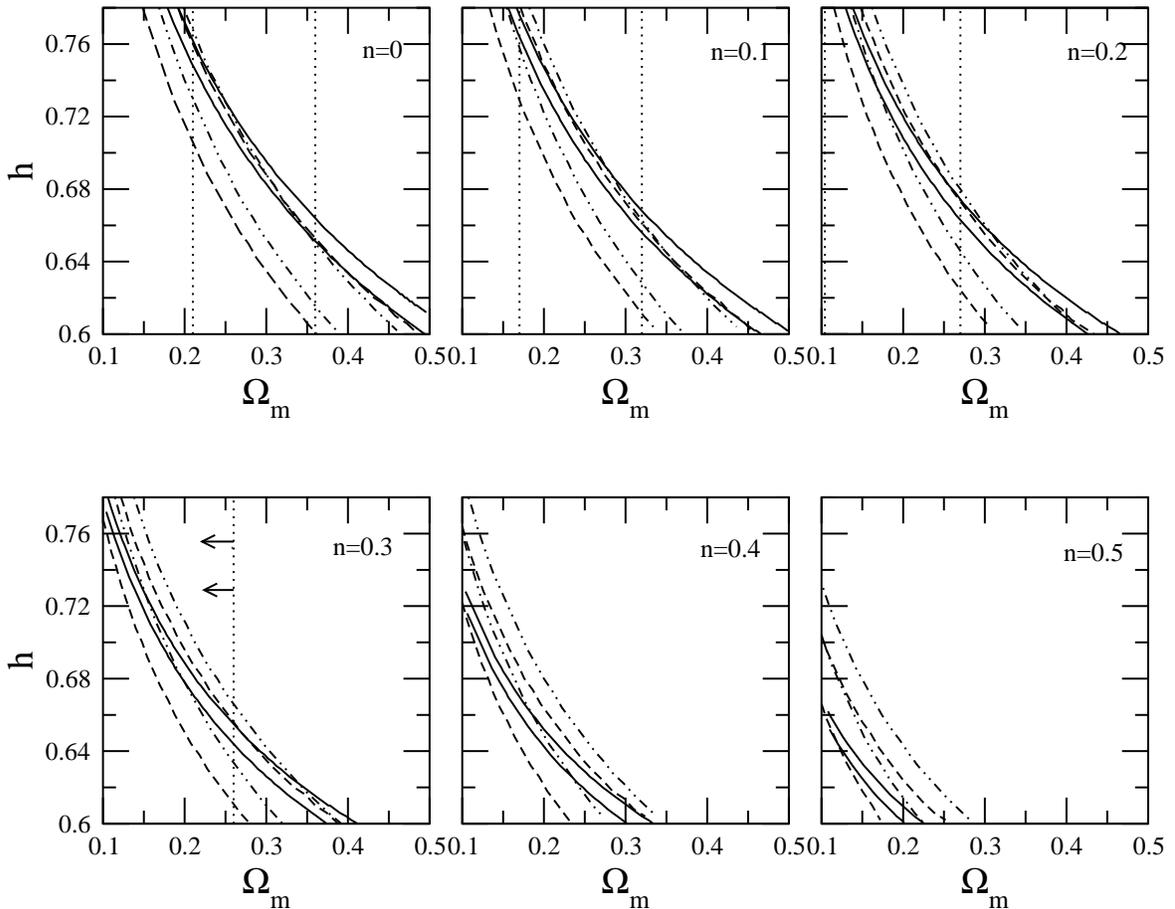}\\
\vskip 0.1cm
\caption{Same as Fig.1 but with $n_s=0.97$.}
\label{fig:figure1}
\end{figure}

To calculate the conformal time today and at the last scattering, we use eqn.(6),
\be 
\tilde\tau_{ls}=\int_0^{\tilde\tau_{ls}} d\tilde\tau=\frac{1}{\Omega_{m0}^{1/2}H_0}\int_0^{a_{ls}}\frac{da}{X(a)} 
\ee
and
\be 
\tilde\tau_{0}=\int_0^{\tilde\tau_{0}} d\tilde\tau=\frac{1}{\Omega_{m0}^{1/2}H_0}\int_0^1 \frac{da}{X(a)} 
\ee
where $X(a)=\sqrt{(a+\frac{\Omega_{r0}}{\Omega_{m0}})+a^{-4n+4}
\left(\frac{1-\Omega_{r0}-\Omega_{m0}}{\Omega_{m0}}\right)
\left(\frac{a+\frac{\Omega_{r0}}{\Omega_{m0}}}{1+\frac{\Omega_{r0}}{\Omega_{m0}}}\right)^n}$.

Substituting the above expression in equation (9), we have the analytical 
expression for $l_A$ in case of this model
\be
l_A=\frac{\pi}{\bar{c_s}}\left[\frac{\int_0^1 \frac{da}{X(a)}}{\int_0^{a_{ls}} \frac{da}{X(a)}}-1\right]
\ee
In an ideal photon-baryon fluid model, the simple analytic relation for the 
position of the m-th peak and the acoustic scale is $l_m=m~l_A$. 
But this simplicity gets disturbed by the different driving and dissipative 
effects which in turn induces a phase shift to the original position\cite{hu}. 
This shift has been compensated by parametrizing the location of the 
peaks and troughs by  
\be    
l_m\equiv l_A(m-\phi_m)\equiv l_A(m-{\bar\phi}-\delta\phi_m)
\ee
where ${\bar\phi}$ is the overall peak shift $\equiv\phi_1$ and 
$\delta\phi_m\equiv\phi_m-{\bar\phi}$ is the relative shift of the mth peak.  
This parametrisation 
can be used to extract information about the matter content of the 
universe before last scattering.  Although it is certainly very difficult 
to derive analytical relation between cosmological parameters and phase 
shifts, Doran and Lilley\cite{doran2} have given certain fitting 
formulae which makes 
life very simple. These formulae does not have a prior and crucially 
depends on cosmological parameters like spectral index $(n_s)$, baryon 
density $(\omega_b=\Omega_b h^2)$, Hubble parameter $(h)$, ratio of radiation 
to matter at last scattering $(r_{ls})$ and also $\Omega^d_{ls}$ as 
representative of dominating present dark energy density at the time of 
recombination. We use these formulae (given in Appendix)to specify 
the positions of the peaks in our model and finally 
constrain the model with the most recent WMAP results.

\begin{figure}[t]
\centering
\leavevmode \epsfysize=6cm \epsfbox{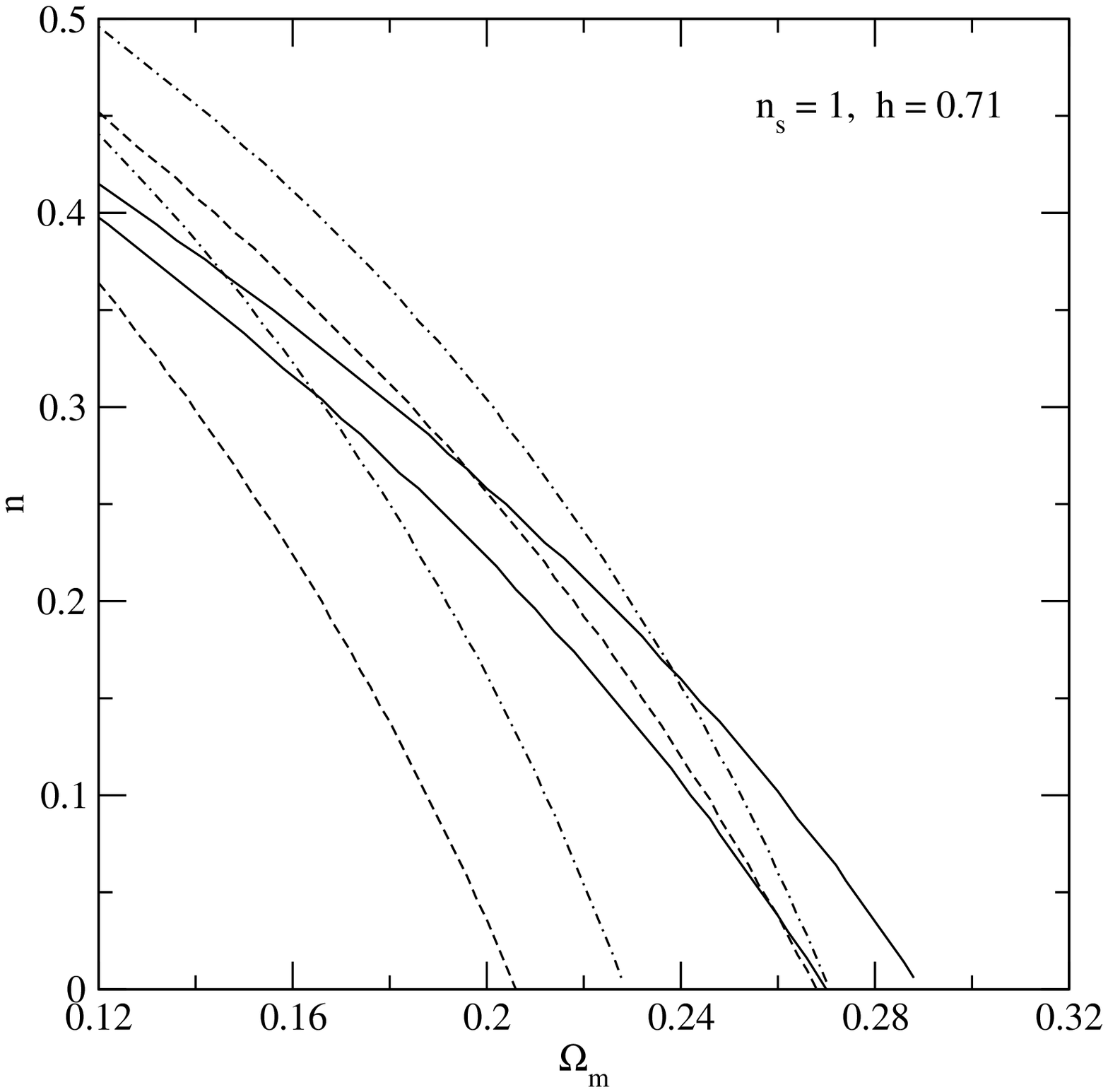}
\hskip 0.6cm 
\leavevmode \epsfysize=6cm \epsfbox{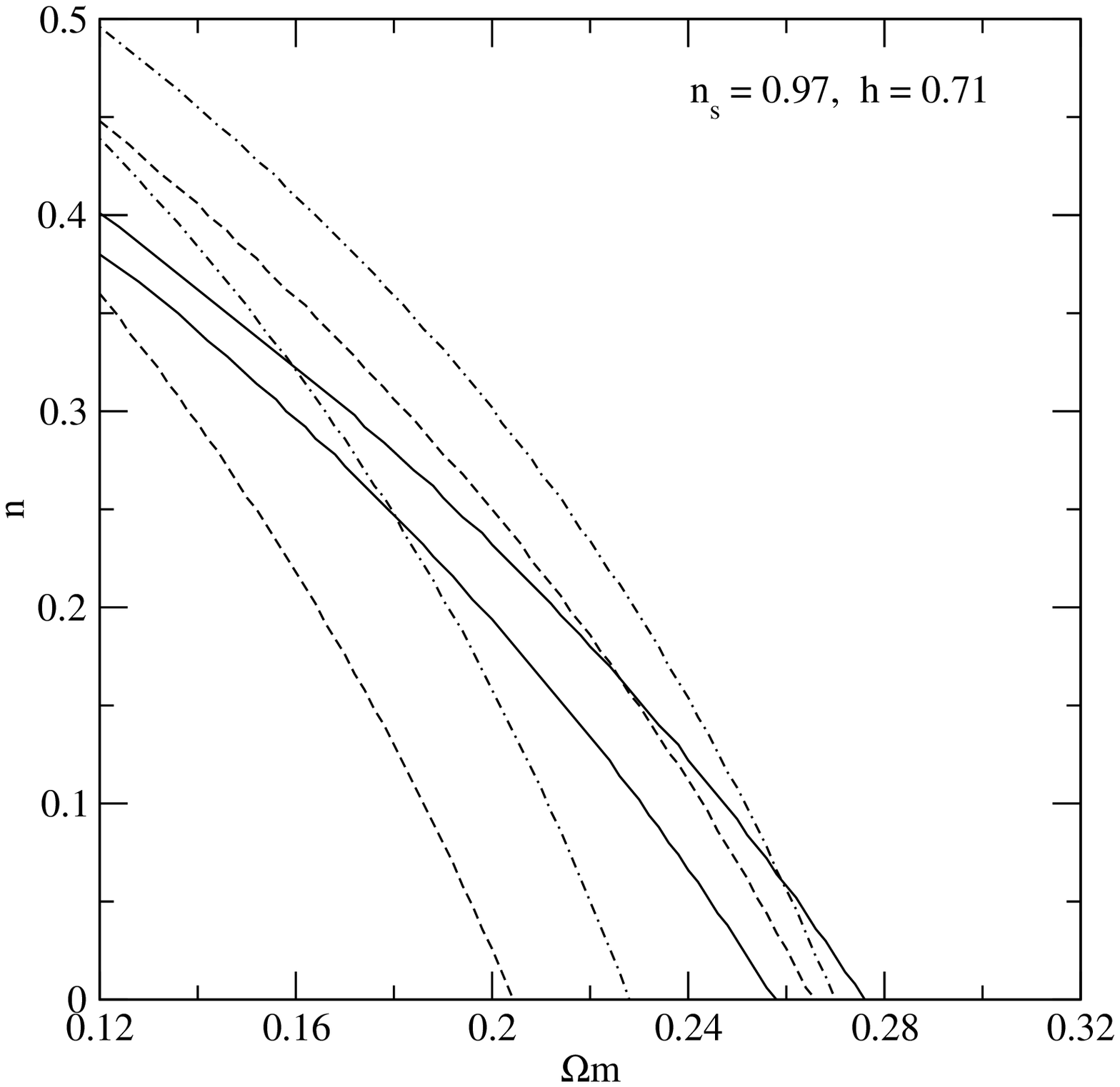}\\
\vskip 0.1cm
\caption{Contour plot of the locations of first, second and third peak locations in ($\Omega_{m},n$) plane for $n_{s}=1,0.97$ and $h=0.71$. The full and dashed lines corresponds to the WMAP constraints on first and second peak locations and the dot-dashed line corresponds to the Boomerang constraint on the third peak location. The allowed region is the intersection of all the contours.}
\label{fig:figure1}
\end{figure}

The locations of the first two acoustic peaks from the WMAP measurement\cite{wmap} of the CMB temperature power 
spectrum are
\bea 
l_{p_1} &=& 220.1\pm 0.8,\nonumber\\
l_{p_2} &=& 546\pm 10;
\eea
notice that all uncertainties are within $1\sigma$. The location for the third peak is given by 
BOOMERanG measurements\cite{boom}
\be
l_{p_3}=825^{+10}_{-13}.
\ee

We have studied the locations of the first three acoustic peaks in the cosmological parameter space for
the Cardassian model given by Eq. (6).The sole parameter of the model is $n$. The cosmological 
parameter space we have investigated is given by $(\tau, n_{s}, h, w_b, \Omega_m)$. Throughout this paper, we have neglected
the contribution from the spatial curvature and massive neutrinos, setting $\Omega_k=\Omega_{\nu}=0$. We have also neglected the 
contribution from the gravitational wave in the initial fluctuation. Because of the rather tight WMAP constraint on $w_b,(w_b=0.0224\pm 0.0009$ \cite{wmap}), we have assumed $w_b=0.0224$ 
in our subsequent calculations. We have also assumed the optical depth of the last scattering 
$\tau=0.11$ which is within the range of the WMAP bound, $\tau=0.166^{+0.076}_{-0.071}$ \cite{wmap}.

In Fig.1, we have plotted the contours of the first three acoustic peak locations, corresponding to the
 WMAP and BOOMERanG bounds given by Eqn.(14)-(15) in the ($\Omega_m,h$) parameter space for different values of $n$ and for 
$n_s=1.0$. In the same figure we have also shown the $1-\sigma$ bound on $\Omega_m$ by fitting our model with the SCP 
(Supernova Cosmology Project) data \cite{sn1}. 

To obtain a fit to SCP data, we write the apparent magnitude as
\be
m(z,\Omega_m,n,{\cal M})=5 log_{10}{\cal D}_l+{\cal M},
\ee
where ${\cal{M}}\equiv M-5\log_{10}H_0+25$ and $D_l=H_0 d_l$ is the 
dimensionless luminosity distance, given in ref \cite{our}. The parameters   
$\Omega_m,n$ and ${\cal M}$ are determined by minimizing 
\be
\chi^2=\sum_i\frac{[m_{exp}(z_i)-m(z_i,\Omega_m,n,{\cal M})]^2}{\sigma^2_i} 
\ee
where ${\sigma^2_i}$ is the error in $m_{exp}(z_i)$. While computing $\Delta\chi^2=\chi^2-\chi^2_{min}$ contours we choose the minimum of ${\cal M}$ at each point of $(\Omega_m,n)$. With a Gaussian distribution, we put the $1-\sigma$ (correspoding to 68.3$\%$) bound on $\Omega_m$ for each set of $n$ across $\chi^2_{min}$.  

In Fig.2, we have plotted the same contours for the CMB peaks with $n_s=0.97$ and Supernova bounds. For both $n_s=1$ and $n_s=0.97$, there is no $1-\sigma$ confidence region of $\Omega_m$ for SCP data within the assumed range (0.1 to 0.4) for $n\geq 0.4$.

In Table 1, we have given the the allowed region for $\Omega_m$ and $h$ coming from WMAP  and BOOMERanG constraints on first three acoustic peak locations with and without the Supernova constraint.

In Fig.3, we have shown the same contours as in Fig.1 and Fig.2, but in ($\Omega_m,n$) plane fixing $h=0.71$. 

\begin{table*}[t]
\caption[par]{\label{tab:par} Limits on $\Omega_m$ and $h$ for different values of $n_s$ and $n$ from CMB constraints}
\begin{center}
\begin{tabular}{|c|c|c|c|c|}
\hline
\multicolumn{5}{|c|}{Without Supernova} \\
\hline
& \multicolumn{2}{|c}{$n_s$ = 0.97} &
\multicolumn{2}{|c|}{$n_s$ = 1.00}  \\
 \cline{2-5} {n} & \multicolumn{1}{|c}{$\Omega_m$} &
\multicolumn{1}{|c|}{$h$} & \multicolumn{1}{|c}{$\Omega_m$} & 
\multicolumn{1}{|c|}{$h$} \\ \hline
\multicolumn{1}{|c|}{0}
& \multicolumn{1}{|c}{$\Omega_m \leq$ 0.35} & \multicolumn{1}{|c|}{ $h\geq$ 0.65} &
\multicolumn{1}{|c}{$\Omega_m \leq$0.25} & \multicolumn{1}{|c|}{$h\geq$0.72}   \\ \hline
\multicolumn{1}{|c|}{0.1}
& \multicolumn{1}{|c}{$\Omega_m \leq$ 0.38} & \multicolumn{1}{|c|}{$h\geq$ 0.63} &
\multicolumn{1}{|c}{$\Omega_m \leq$0.29} & \multicolumn{1}{|c|}{$h\geq$0.68}   \\ \hline
\multicolumn{1}{|c|}{0.2}
& \multicolumn{1}{|c}{0.13$\leq\Omega_m \leq$ 0.40} & \multicolumn{1}{|c|}{0.60$\leq h\leq$ 0.78} &
\multicolumn{1}{|c}{$\Omega_m \leq$0.32} & \multicolumn{1}{|c|}{$h\geq$0.64}   \\
\hline 
\multicolumn{1}{|c|}{0.3}
& \multicolumn{1}{|c}{$\Omega_m \geq$ 0.15} &
\multicolumn{1}{|c|}{$h\leq$0.72} & \multicolumn{1}{|c}{0.13$\leq\Omega_m \leq$0.35} & \multicolumn{1}{|c|}{0.76$\geq h\geq$0.61}   \\ \hline
\multicolumn{1}{|c|}{0.4}
& \multicolumn{1}{|c}{$\Omega_m \geq$ 0.18} &
\multicolumn{1}{|c|}{$h\leq$0.66} & \multicolumn{1}{|c}{$\Omega_m \geq$0.16} & \multicolumn{1}{|c|}{$h\leq$0.68}   \\ \hline
\multicolumn{1}{|c|}{0.5} & \multicolumn{1}{|c}{-} &
\multicolumn{1}{|c|}{-} &
\multicolumn{1}{|c}{$\Omega_m \geq$0.19} & \multicolumn{1}{|c|}{$h\leq$0.62}   \\ \hline
\end{tabular}
\begin{tabular}{|c|c|c|c|c|}
\hline
\multicolumn{5}{|c|}{With Supernova} \\
\hline
& \multicolumn{2}{|c}{$n_s$ = 0.97} &
\multicolumn{2}{|c|}{$n_s$ = 1.00}  \\
 \cline{2-5} {n} & \multicolumn{1}{|c}{$\Omega_m$} &
\multicolumn{1}{|c|}{$h$} & \multicolumn{1}{|c}{$\Omega_m$} 
& \multicolumn{1}{|c|}{$h$} \\ \hline
\multicolumn{1}{|c|}{0}
& \multicolumn{1}{|c}{0.21$\leq\Omega_m \leq$0.35} & \multicolumn{1}{|c|}{0.65$\leq h\leq$0.76} &
\multicolumn{1}{|c}{0.21$\leq\Omega_m \leq$0.25} & \multicolumn{1}{|c|}{0.72$\leq h\leq$0.77}   \\ \hline
\multicolumn{1}{|c|}{0.1}
& \multicolumn{1}{|c}{0.17 $\leq\Omega_m \leq$0.32} & \multicolumn{1}{|c|}{0.66$\leq h\leq$ 0.77} &
\multicolumn{1}{|c}{0.17$\leq\Omega_m \leq$ 0.29} & \multicolumn{1}{|c|}{0.68$\leq h\leq$ 0.78}   \\ \hline
\multicolumn{1}{|c|}{0.2}
& \multicolumn{1}{|c}{0.13$\leq\Omega_m \leq$ 0.27} & \multicolumn{1}{|c|}{0.67$\leq h\leq$ 0.78} &
\multicolumn{1}{|c}{$\Omega_m \leq$ 0.27} & \multicolumn{1}{|c|}{$h\geq$ 0.67}   \\
\hline 
\multicolumn{1}{|c|}{0.3}
& \multicolumn{1}{|c}{0.15$\leq\Omega_m \leq$ 0.26} &
\multicolumn{1}{|c|}{0.65$\leq h\leq$ 0.72} & \multicolumn{1}{|c}{0.13$\leq\Omega_m \leq$ 0.26} & \multicolumn{1}{|c|}{0.66$\leq h\leq$ 0.76}   \\ \hline
\end{tabular}
\end{center}
\end{table*}

\section{Constraints on $\sigma_8$}

In this section we present the WMAP constraints on $\sa$ which is related to the amplitude of the galaxy fluctuations for different values of the model parameter $n$. $\sa$, the rms density fluctuations averaged over $8h^{-1}$ Mpc spheres, is determined by the COBE normalisation of the CMB power spectrum. This is done by essentially fixing the fluctuations at the last scattering for a given model. 

Doran et.al\cite{doran3} have given an estimate of the CMB-normalized 
$\sa$-value for 
a very general class of 
dark energy models just from the knowledge of  their ``background solution'' 
$[\omd(a),\ \wda(a)]$ (here superscript ``d'' stands for dark energy) 
and the $\sa$-value of the $\Lambda$CDM model, with 
 $\Omega^{\Lambda}_0=\Omega^d_0(\Lambda)$:
\begin{equation} \label{main}
 \frac{\sigma _8 (d)}{\sigma _8 (\Lambda)}\approx
\left( a _{\rm eq}\right)^{ 3\, \omdsf / 5}
 \left(1-\Omega^{\Lambda}_0 \right)^{-\left (1+ \bar w ^{-1}\right)/5}
 \sqrt{\frac{\tilde\tau _0 (d)}{\tilde\tau _0 (\Lambda)}}.
\end{equation} 
where $\tilde\tau _0$ is the conformal age of 
the universe, $\aeq$ is the scale factor at matter radiation equality
given by,
\begin{equation}\label{equality}
  \aeq=\frac{\Omega _{r0}}{\Omega _{m0}}= \frac{4.31 \times 10^{-5}}
  {h^2(1-\omdn)}.
\end{equation}
and $\bar{\omega}$ is the effective equation of state which is an
average value for $\wda$ during the time in which $\omd$ is growing rapidly:
\begin{equation}
\frac{1}{\bar{\omega}}=\frac{\int _{\ln a_{\rm tr}}^0 \omd(a)/\wda(a)\:  d \ln{a}}
                   {\int _{\ln a_{\rm tr}}^0 \omd(a)\:  d  \ln{a}}.
\end{equation}
As shown in one of our previous papers \cite{our}, the extra term in equation 1 can be represented as dark energy component with slowly varying equation of state and in that case ${\wda}_0$ can be used as a fair approximation to $\bar{\omega}$. In our case it is $n-1$\cite{our}.

$\omdsf$ is an average value for the fraction of dark energy 
during the matter dominated era, before $\omd$ starts growing rapidly 
\begin{equation}
 \omdsf \equiv [ \ln{a_{\rm tr}}-\ln{\aeq} ]^{-1} \int_{\ln \aeq}^{\ln a_{\rm tr}} \omd(a)\  {\rm d} \ln a  .
\end{equation}
For model without early quintessence, which is precisely our case, $\omdsf$ is zero. 

In order to compute $\sa$ for the $\Lambda$CDM model, we have used the standard definition
\be
\sa^2 \equiv \int^\infty_0 {d\kappa\over{\kappa}}\Delta^2(\kappa)\left({3j_1(\kappa r)\over{\kappa r}}\right)^2,
\ee
with $r=8h^{-1}$ Mpc and
\be
\Delta^2(\kappa) = \delta^2_H \left({\kappa\over{H_0}}\right)^{3+n_s} T^2(\kappa),
\ee
$T(\kappa)$ is the matter transfer function describing the processing of the initial fluctuations, for which we use the result \cite{transf}
\be
T(q) = {\ln(1+2.34q)\over{2.34q}}[1+3.89q+(16.1q)^2+(5.46q)^3+(6.71q)^4]^{-1/4},
\ee
with $q=\frac{\kappa}{h\Gamma}Mpc$ and $\Gamma$ is \cite{transf}
\be
\Gamma = \Omega_m h \exp\left[-\Omega_b\left(1+{\sqrt{2h}\over{\Omega_m}}\right)\right].
\ee

$\delta_H$ is the density perturbation at horizon crossing and a fit to the four-year COBE data gives \cite{delh}
\be
10^5 \delta_H (n_s,\Omega_m)=1.94 \Omega_m^{-0.785-0.05\ln\Omega_m}\exp[-0.95(n_s-1)-0.169(n_s-1)^2]
\ee
assuming that there is no gravitational waves and also no reionization. Taking in to account the reionization effects, the above expression has been corrected by Griffiths and Liddle \cite{reion}:

\be
{\delta_H(\tau)\over{\delta_H(\tau=0)}}=1+0.76\tau-1.96\tau^2+1.46\tau^3,
\ee

where $\tau$ is the optical depth. The above correction is reliable upto $\tau=0.5$. As mentioned earlier we have used $\tau=0.11$ in our calculation.

In Table II, we show the constraint on $\sa$ for different values of $n$ and $n_s$, normalizing the calculation at $\Omega_m=0.25$.

\begin{table*}[t]
\caption[par2]{\label{tab:par2} Limits on $\sigma_8$ for different values of $n_s$ and $n$ and $\Omega_m=0.25$}
\begin{center}
\begin{tabular}{|c|c|c|}
\hline
\multicolumn{3}{|c|}{With Supernova} \\
\hline
& \multicolumn{1}{|c}{$n_s$ = 1} &
\multicolumn{1}{|c|}{$n_s$ = 0.97}  \\
 \cline{2-3} {n} & \multicolumn{1}{|c}{$\sigma_8$}  & \multicolumn{1}{|c|}{$\sigma_8$} \\ \hline
\multicolumn{1}{|c|}{0}
& \multicolumn{1}{|c}{0.83$\leq\sigma_8\leq$0.85}&
\multicolumn{1}{|c|}{0.77$\leq\sigma_8\leq$0.79} \\ \hline
\multicolumn{1}{|c|}{0.1}
& \multicolumn{1}{|c}{0.78$\leq\sigma_8\leq$0.80} &
\multicolumn{1}{|c|}{0.72$\leq\sigma_8\leq$ 0.74} \\ \hline
\multicolumn{1}{|c|}{0.2}
& \multicolumn{1}{|c}{0.73$\leq\sigma_8\leq$ 0.75} & 
\multicolumn{1}{|c|}{0.67$\leq\sigma_8\leq$ 0.69}  \\
\hline 
\multicolumn{1}{|c|}{0.3}
& \multicolumn{1}{|c}{0.67$\leq\sigma_8\leq$ 0.68} & \multicolumn{1}{|c|}{0.61$\leq\sigma_8\leq$ 0.63}\\ \hline
\end{tabular}

\end{center}
\end{table*}

\section{Discussion}

The first result is that the WMAP constraint on the first two acoustic peak locations together with the BOOMERanG bound on the third peak location restricts the model parameter $n$ as $n\leq0.5$ for $n_s=1$ and $n\leq0.4$ for $n_s=0.97$.
One should note that to have current acceleration in this model, $n\leq 0.66$ is required, which means that CMB data alone predicts a late time accelerating universe. When one puts the Supernova constraint on top of this, the bound on $n$ becomes $n\leq0.3$ for both values of $n_s$.  Another interesting feature of our analysis is that in order to distinguish between  a $\Lambda$CDM model and a Cardassian model, one has to impose a low value of $h$ (e.g. $h\leq0.71$) and higher value of $n_s$ ($n_s \approx 1$). This has also been shown in Fig.3 where we have plotted the contours of the peak locations in the ($\Omega_m,n$) plane fixing $h=0.71$. It is clear that for $n_s=1$, the allowed region does not include $n=0$ which is precisely the $\Lambda$CDM case. But for $n_s=0.97$ it includes $n=0$. In other words, for low values of $h$ together with $n_s$ closer to $1$, it appears that the $\Lambda$CDM model is more disfavoured than the Cardassian model. On the other hand for smaller values of $n_s$, $\Lambda$CDM and Cardassian model are equally favoured. An interesting point to note here is that WMAP Constraint on $n_s$ is $n_s=0.99\pm0.04$. Hence although this remains still an unresolved issue, the fact that $n_s$ plays an important role in determining the nature of dark energy is one very interesting outcome of this investigation. We have shown it in the context of Cardassian model (the fact, that $n_s$ plays an important role in determining the nature of dark energy, has also been shown recently by Barreiro et.al \cite{bbss} in the context of early quintessence model), but it would be very interesting to study this feature in a model independent way. 

We have also calculated the WMAP bound on $\sa$ in our model for different values of $n$ and $n_s$ normalizing the calculation at $\Omega_m=0.25$. Table II shows that $\sa$ gets lower for higher value of $n$ and lower value of $n_s$. Considering the fact that the joint WMAP/Large scale structure data set indicate  a suppressed clustering power at small scale \cite{wmap},resulting a lower $\sa$ value, one can conclude  that lower value of $n_s$ and higher value of $n$ is favoured so far structure formation is concerned.

The greatest challenge in cosmology today is to unravel the nature of dark energy. Although the WMAP observations has put stringent constraint on many cosmological parameters, but there remains still some uncertainty in some important parameters like $n_s$, $\sa$, $\Omega_m$ which plays very crucial roles in  determining the nature of the dark energy.  Hence the future observations with stronger constraints on these parameters hold promise to uncover the nature of dark energy. 
\acknowledgments
The work of A.A.S. is fully 
financed by  Funda\c c\~ao para a Ci\^encia e a Tecnologia (Portugal)
under the grant POCTI/1999/FIS/36285. The work of S.S. is financed by 
 Funda\c c\~ao para a Ci\^encia e a Tecnologia (Portugal), through CAAUL.

\section{appendix}

For completeness, we repeat here the formulas used in our search for parameter space.
These fitting formulae are quoted from the cited literature\cite{doran2}.
 
We assume the standard recombination history and define the redshift of 
decoupling
$z_{ls}$ as the redshift at which the optical depth of Thompson scattering 
is unity.
A useful fitting formula for $z_{ls}$ is given by \cite{param}:
\be
z_{ls}=1048[1+0.00124w_b^{-0.738}][1+g_1w_m^{g_2}],
\label{zdec}
\ee
where
$$
g_1=0.0783w_b^{-0.238}[1+39.5w_b^{0.763}]^{-1},\;\;\;
g_2=0.56[1+21.1w_b^{1.81}]^{-1},
$$
$w_b\equiv\Omega_bh^2$ and $w_m\equiv\Omega_mh^2$.

And the ratio of radiation to matter at last
scattering 
\begin{equation}\label{r_star}
r_{ls} = \rho_r(z_{ls}) / \rho_m(z_{ls}) = 0.0416w_m^{-1} \left(z_{ls} / 10^3\right). 
\end{equation}

The overall phase shift $\bar{\varphi}$ (which is basically the phase shift of
the first peak) is parametrised by the following formula
\begin{equation} \label{phi_1}
\bar{\varphi} =(1.466 - 0.466n_s) \left[ a_1 r_{{ls}}^{a_2} + 0.291
  \bar\Omega_{ls}^{\phi} \right],
\end{equation}
where $a_1$ and $a_2$ are given by
\begin{eqnarray}
a_1 & = & 0.286 + 0.626~w_b \\
a_2 & = & 0.1786 - 6.308~w_b + 174.9~w_b^2 - 1168~w_b^3.
\end{eqnarray}
and $\bar\Omega_{ls}^d$ is the average fraction of dark energy before last  scattering which is zero for our case.

The relative shift of the second peak $(\delta\phi_2)$ is given by
\begin{equation}
\delta \varphi_2 = c_0 - c_1 r_{ls} - c_2 r_{ls} ^{-c_3} + 0.05\,
(n_s-1),
\end{equation}
with
\begin{eqnarray}
c_0 &=& -0.1 + \left( 0.213 - 0.123 \bar\Omega_{ls}^d \right)\\
&&\times \exp\left\{ - \left( 52 - 63.6\bar\Omega_{ls}^d\right)
w_b \right\}\\
c_1 &=& 0.063 \,\exp \{-3500~w_b^2 \} +
0.015\\
c_2 &=& 6\times 10^{-6} + 0.137 \left (w_b - 0.07 \right) ^2\\ 
c_3 &=& 0.8 + 2.3 \bar\Omega_{ls}^d + \left( 70 - 126\bar\Omega_{ls}^d\right) w_b.
\end{eqnarray}

For the third peak, 
\begin{equation}
\delta \varphi_3 = 10 - d_1r_{ls}^{d_2} + 0.08\, (n_s-1),
\end{equation}
with 
\begin{eqnarray}
d_1 &=& 9.97 + \left(3.3 -3 \bar\Omega_{ls}^d\right)w_b \\
\nonumber d_2 &=& 0.0016 - 0.0067\bar\Omega_{ls}^d + \left(0.196 - 0.22\bar\Omega_{ls}^d\right)w_b\\ 
&& + \frac{(2.25 + 2.77 \bar\Omega_{ls}^d) \times 10^{-5}}{w_b},
\end{eqnarray}

The over all shifts for the second and the third peak is 
$\bar\phi+\delta\phi_2$ and $\bar\phi+\delta\phi_3$ respectively. 



\begin{thebibliography}{99}

\bibitem{wmap}D.N.Spergel et al., astro-ph/0302207;L.Page et al, astro-ph/0303220
\bibitem{sn1}S. Perlmutter, M. Della Valle et al., Nature, {\bf 391},
 (1998); S. Perlmutter, G. Aldering, G. Goldhaber et al., Astroph. J.,
 {\bf 517} (1999)
\bibitem{sn2}P.M. Garnavich, R.P. Kirshner, P. Challis et al., Astroph
. J., {\bf 493}, (1998); A.G. Riess, A.V. Philipenko, P. Challis et al.,
Astron. J., {\bf 116}, 1009 (1998).
\bibitem{2df}M.Colless et al., MNRAS, {\bf 328},1039,(2001); W.Percival, MNRAS, {\bf 327}, 1297 (2001).
\bibitem{cmb}P. Bernadis et al., Nature, {\bf 404}, 955 (2000);
S. Hanany et al., astro-ph/0005123; A. Balbi et al., 
astro-ph/0005124.
\bibitem{quint}R.R. Caldwell, R. Dave and P.J. Steinhardt,
Phys. Rev.
Lett, {\bf 80}, 1582 (1998);
P.J.E. Peebles and B. Ratra,
Astrophys.J.Lett., {\bf
325}, L17, (1988); P.G. Ferreira and M. Joyce,
Phys.Rev.Lett., {\bf
79}, 4740 (1987); E.J. Copeland, A.R. Liddle and
D. Wands,
Phys.Rev.D, {\bf 57}, 4686 (1988);
P.J. Steinhardt, L. Wang and I. Zlatev,
Phys.Rev.Lett., {\bf 59},
123504 (1999);
I. Zlatev, L. Wang and P.J. Steinhardt,
Phys.Rev.Lett.,
{\bf 82}, 896 (1999);
C. Wetterich, Nucl.Phys.B {\bf 302}, 668, (1988);
B. Ratra and P.J.E. Peebles Phys.Rev.D, {\bf 37}, 3406 (1988);
T. Barreiro, E.J. Copeland and N.J. Nunes, Phys.Rev.D, {\bf 61},
127301 (2000); 
V. Sahni and L. Wang, Phys.Rev.D, {\bf 59}, 103517 (2000);
A.A. Sen and S. Sethi, Phys.Lett.B, {\bf 532}, 159 (2002);
M.C. Bento, O. Bertolami, N.C. Santos astro-ph/0106405.
\bibitem{ksen} C. Armendariz-Picon, V. Mukhanov and  P.J. Steinhardt, Phys.Rev.Lett, {\bf 85}, 4438 (2000);
T. Chiba, Phys.Rev.D, {\bf 66}, 063514 (2002).
\bibitem{others}N. Bertolo and M. Pietroni, Phys.Rev.D, {\bf 61}, 023518 (1999);
O. Bertolami and P.J. Martins, Phys.Rev.D, {\bf 61}, 064007
(2000);
J.P. Uzan Phys.Rev.D, {\bf 59}, 123510 (1999);
N. Banerjee and D. Pavon, Phys.Rev.D, {\bf 63}, 043504 (2001);
Class.Quant.Grav, {\bf 18}, 593 (2001);
S. Sen and T. R. Seshadri gr-qc/0007079;
L. Amendola, Phys.Rev.D, {\bf 62}, 043511 (2000);
L. Amendola, Phys.Rev.D, {\bf 60}, 043501 (1999);
T. Chiba, Phys.Rev.D, {\bf 60}, 083508 (1999);
M. Gasperini, gr-qc/0105082;
M. Gasperini, F. Piazza and G. Veneziano, gr-qc/0108016;
A. Riazuelo and J. Uzan, astro-ph/0107386;
A.A. Sen, S. Sen and S. Sethi, Phys.Rev.D, {\bf 63}, 107501 (2001);
S. Sen and A.A. Sen, Phys.Rev.D, {\bf 63}, 124006 (2001);
A.A. Sen and S. Sen, Mod.Phys.Lett.A, {\bf 16}, 1303, (2001).
\bibitem{chap}A. Kamenshchik, U. Moschella, V. Pasquier, Phys.Lett.B  {\bf 511}, 265 (2001);
M.C. Bento, O. Bertolami and A.A. Sen, Phys.Rev.D, {\bf D66} 043507 (2002);
N. Bili\'c, G.B. Tupper, R.D. Viollier, Phys.Lett.B,  {\bf 535}, 17
 (2002);
M.C. Bento, O. Bertolami and A.A. Sen, astro-ph/0210468.
\bibitem{paddy}T. Padmanabhan and T. Roy Choudhury, Phys.Rev.D, {\bf 66}, 081301(R) (2002).
\bibitem{freese}K. Freese and M. Lewis, Phys.Lett.B, {\bf 540}, 1 (2002).
\bibitem{motiv}D.J. Chung and K. Freese, Phys.Rev.D, {\bf 61}, 023511 (2000);
P. Gondolo and K. Freese, hep-ph/0209322; 
P. Gondolo and K. Freese, hep-ph/0211397
\bibitem{our}S. Sen and A.A. Sen, astro-ph/0211634, to appear in Astrophys. J.
\bibitem{other}Zhu,Z. and Fujimoto,M., Astrophys. J, {\bf 581}, 1 (202);
Y. Wang, K. Freese, P. Gondolo and M. Lewis, astro-ph/0302064.
\bibitem{hu}W. Hu, M. Fukugita, M. Zaldarriaga and M. Tegmark, Astrophys.J, {\bf 549}, 669 (2001).
\bibitem{doran1}M. Doran, M. Lilley, J. Schwindt and C. Wetterich, Astrophys.J,{\bf 559}, 501(2001).
\bibitem{doran2}M. Doran and M. Lilley, Mon.Not.Roy.Astron.Soc., {\bf 330} 965 (2002).
\bibitem{boom}J.E.Ruhl et al,astro-ph/0212229. 
\bibitem{doran3}M. Doran, J.M. Schwindt and C. Wetterich, Phys.Rev.D, {\bf 64},123520, (2001).
\bibitem{transf}J.M. Bardeen, J.R. Bond, N. Kaiser and A.S. Szalay, Astrophys. j. {\bf 304}, 15 (1986).
\bibitem{delh}E.F.Bunn and M.J.White, Astrophys. J., {\bf 480}, 6, (1997).
\bibitem{reion}L.M.Griffiths and A.R.Liddle, astro-ph/0101149.
\bibitem{param}R.Durrer, D. Novosyadlyj and S.Apunevych, astrophys.J., {\bf 583}, 33, (2003).
\bibitem{bbss}T. Barreiro, M.C. Bento, N.M.C. Santos and A.A. Sen, astro-ph/0303298.
\end{thebibliography}
\end{document}